\documentclass[prl,superscriptaddress,twocolumn,showpacs,preprintnumbers,amsmath,amssymb]{revtex4}

\usepackage{graphicx}
\usepackage{dcolumn}
\usepackage{bm}

\begin{document}


\title{ $\Lambda d$ correlations from the $^4$He(stopped-$K^-$,$d$) reaction}

\author{T. Suzuki}
\email{takatosi@riken.jp}
\affiliation{RIKEN Nishina Center, RIKEN, Saitama 351-0198, Japan \\}
\author{H. Bhang}
\affiliation{Department of Physics, Seoul National University, Seoul 151-742, South Korea \\}
\author{J. Chiba}
\affiliation{ Department of Physics, Tokyo University of Science, Chiba 278-8510, Japan \\}
\author{S. Choi}
\affiliation{Department of Physics, Seoul National University, Seoul 151-742, South Korea \\}
\author{Y. Fukuda}
\affiliation{Department of Physics, Tokyo Institute of Technology, Tokyo 151-8551, Japan \\}
\author{T. Hanaki}
\affiliation{ Department of Physics, Tokyo University of Science, Chiba 278-8510, Japan \\}
\author{R. S. Hayano}
\affiliation{Department of Physics, University of Tokyo, Tokyo 113-0033, Japan \\}
\author{\\M. Iio}
\affiliation{RIKEN Nishina Center, RIKEN, Saitama 351-0198, Japan \\}
\author{T. Ishikawa}
\affiliation{Department of Physics, University of Tokyo, Tokyo 113-0033, Japan \\}
\author{ S. Ishimoto}
\affiliation{High Energy Accelerator Research Organization (KEK), Ibaraki 305-0801, Japan \\}
\author{ T. Ishiwatari}
\affiliation{Stefan Meyer Institut f\"{u}r subatomare Physik, A-1090 Vienna, Austria \\}
\author{ K. Itahashi}
\affiliation{RIKEN Nishina Center, RIKEN, Saitama 351-0198, Japan \\}
\author{ M. Iwai}
\affiliation{High Energy Accelerator Research Organization (KEK), Ibaraki 305-0801, Japan \\}
\author{M. Iwasaki}
\affiliation{RIKEN Nishina Center, RIKEN, Saitama 351-0198, Japan \\}
\affiliation{Department of Physics, Tokyo Institute of Technology, Tokyo 151-8551, Japan \\}
\author{\\P. Kienle}
\affiliation{Stefan Meyer Institut f\"{u}r subatomare Physik, A-1090 Vienna, Austria \\}
\affiliation{Physik Department, Technische Universit\"{a}t M\"{u}nchen, D-85748 Garching, Germany \\}
\author{J. H. Kim}
\altaffiliation[Present address: ]{Korea  Research Institute of Standards and Science, Taejon 305-340, South Korea \\}
\affiliation{Department of Physics, Seoul National University, Seoul 151-742, South Korea \\}
\author{Y. Matsuda}
\affiliation{RIKEN Nishina Center, RIKEN, Saitama 351-0198, Japan \\}
\author{H. Ohnishi}
\affiliation{RIKEN Nishina Center, RIKEN, Saitama 351-0198, Japan \\}
\author{S. Okada}
\affiliation{RIKEN Nishina Center, RIKEN, Saitama 351-0198, Japan \\}
\author{H. Outa}
\affiliation{RIKEN Nishina Center, RIKEN, Saitama 351-0198, Japan \\}
\author{M. Sato}
\altaffiliation[Present address: ]{RIKEN Nishina Center, RIKEN, Saitama 351-0198, Japan \\}
\affiliation{Department of Physics, Tokyo Institute of Technology, Tokyo 151-8551, Japan \\}
\author{\\S. Suzuki}
\affiliation{High Energy Accelerator Research Organization (KEK), Ibaraki 305-0801, Japan \\}
\author{D. Tomono}
\affiliation{RIKEN Nishina Center, RIKEN, Saitama 351-0198, Japan \\}
\author{E. Widmann}
\affiliation{Stefan Meyer Institut f\"{u}r subatomare Physik, A-1090 Vienna, Austria \\}
\author{T. Yamazaki}
\affiliation{Department of Physics, University of Tokyo, Tokyo 113-0033, Japan \\}
\affiliation{RIKEN Nishina Center, RIKEN, Saitama 351-0198, Japan \\}
\author{H. Yim}
\affiliation{Department of Physics, Seoul National University, Seoul 151-742, South Korea \\}

\collaboration{KEK-PS E549 collaboration}
\date{\today}

\begin{abstract}
We have observed an intense high-energy component in an inclusive $^4$He(stopped $K^-$, $d$) spectrum. For back-to-back $\Lambda d$ pairs a prominent event cluster has been found just below the $m_{^4\textrm{He}}+m_{K^-}-m_{n}$ mass threshold in the $\Lambda d$ invariant mass spectrum of $\Lambda dn$ events, which is evidence for a three-nucleon absorption process of $K^-$  in $^4$He. In addition, an appreciable strength is revealed below $\sim$3220 MeV/$c^2$. Well separated $\Sigma^0 dn$ events show a peak similar to the case of $\Lambda dn$. 
\end{abstract}

\maketitle

Motivated by the prediction of strongly bound $\bar{K}$-nuclear states 
in several light nuclei~\cite{AY}, an experimental search for $K^-ppn$/$K^-pnn$
states via $^4$He(stopped $K^-$, $N$) reaction was performed at the KEK 12 GeV 
Proton Synchrotron~(KEK-PS E471)~\cite{Iwa2}. As a result, the E471 collaboration reported on the observation and indication of strange tribaryonic states, S$^0$(3115) from a proton spectrum~\cite{TS1} and S$^+$(3140) from  a neutron spectrum~\cite{Iwa1}. Later, the FINUDA collaboration reported on evidence for a strongly-bound $K^-pp$ state~\cite{AY2} from a $\Lambda p$ invariant mass spectrum using stopped $K^-$ reactions on several light nuclear targets~\cite{FINUDA}. These di- and tri-baryonic states triggered controversial discussion of their existence and nature~\cite{Weise}. The states reported in the stopped $K^-$ experiments were discussed in relation to their non-mesonic decay modes, and thus, discrimination against non-mesonic multibody absorption processes are vitally important, especially if the states are relatively broad. These multibody absorption processes of $K^-$ are of the form,
 \begin{eqnarray}
 K^- ``NN"(NN) &\rightarrow& YN(NN) :\textrm{2NA},\\
 K^- ``NNN"(N)& \rightarrow& YNN(N) :\textrm{3NA},\\
 K^- ``NNNN" &\rightarrow& YNNN \thickspace \thickspace \medspace :\textrm{4NA},
 \end{eqnarray} 
 where $``NN"$, $``NNN"$ and $``NNNN"$ denote bound nucleons participating in the $K^-$ absorption, $(NN)$ and $(N)$ stand for spectators ($NN$ includes $d$). According to the number of participating nucleons, we call them as 2-, 3-, and 4-Nucleon Absorption (2, 3, and 4NA) processes. In contrast to the case of pion absorption~\cite{3N,4N} the $K^-$ absorption provides a unique playground in studying such multinucleon absorption processes, since the leading participant can be clearly identified as a well resolved hyperon and large energies and momenta are imparted to other participants. However, the presently available information from bubble chamber experiments is mainly on the total multinucleonic capture rate of stopped $K^-$~\cite{Katz}. Thus, it is important to obtain more information on these reaction branches from (stopped $K^-$, $N/d$) reactions not only for the experimental search of strange mutibaryonic systems, but also for the unified understanding of the low-energy $\bar{K}$-nucleus interaction, for which the multi-nucleon absorption processes are the primary source of the imaginary potential in the deeply bound region where $\bar{K}N\rightarrow Y\pi$ channels are suppressed.

In order to confirm the existence of S$^0$/S$^+$ tribaryons and to search for other possible multibaryonic states, we measured $^4$He(stopped $K^-$, $N/d$) spectra in the KEK-PS E549 experiment. In the deuteron channel, the charge-neutral strange dibaryon with isospin $T=1/2$, which could be an isobaric analogue state of bound nuclear $K^-pp$ system~\cite{AY2,FINUDA}, may appear via the reaction

\begin{equation}
(K^- {}^4\textrm{He})_{\textrm{atomic}} \rightarrow {}^2\textrm{S}^{0}_{T=1/2}+d,
\end{equation}
where we generally denote multibaryonic states with strangeness -1, baryon number $A$ and charge $Z$ as ${}^A$S$^Z$. Furthermore,  the tribaryon production and decay processes, 
\begin{equation}
(K^- {}^4\textrm{He})_{\textrm{atomic}} \rightarrow {}^3\textrm{S}^{+}_{T=0}+n,\atop
{}^3\textrm{S}^{+}_{T=0} \rightarrow \Lambda+d,
\end{equation}
 can be investigated by invariant mass spectroscopy of $\Lambda$ and $d$. In this Letter, we report on results of the deuteron inclusive and $\Lambda$-$d$ coincidence measurements.
 
 The experiment E549 was performed at the K5 beamline at the KEK 12 GeV PS as one of its last experimental programs, with an upgraded experimental apparatus, better resolution and one order of magnitude larger statistics compared to those of the E471. Detailed descriptions of the setup and analysis procedures for specific channels are given in  Refs.~\cite{Sato,Yim}.  
  In the experiment, all particle species were measured by the Time-of-Flight method (TOF). 
 Two high-performance TOF walls, which were symmetrically placed at the left and right side of the target with their front faces positioned at 1.8 m from the center, enabled us to measure $p$, $n$, $\pi^{\pm}$ and $d$ with comparable resolutions of $\Delta(1/\beta)=0.02 \sim 0.03$ in ``inclusive" condition, and they further allowed the coincidence measurement of ``back-to-back" $dp$ pairs emitted horizontally. Two additional detector arms, which were dedicated to measure particles emitted vertically, were installed on the top and bottom of the target, and charged or neutral particles were detected in coincidence with the particles on the TOF walls. This condition is called ``vertical-$X$-coincidence" condition, where $X$ could be $\pi^{\pm},p,n$ and $\gamma$.
   Only incident $K^-$ and an outgoing charged particle were required at the hardware trigger level to measure inclusive proton and deuteron spectra, and particle coincidences were obtained by analysis from the accumulated data. All charged particles were tracked back with drift chambers to the reaction vertex, corrected for energy loss, and $\Lambda$ and $\Sigma^{\pm}$ hyperons were reconstructed by means of $N\pi$ invariant mass spectroscopy. For the present analysis, $\sim 2.5\times10^8$ stopped $K^-$ on the $^4$He target were effectively used, and $1.12 \times 10^5$ deuterons have been identified.
 
\begin{figure}
\includegraphics[scale=.35]{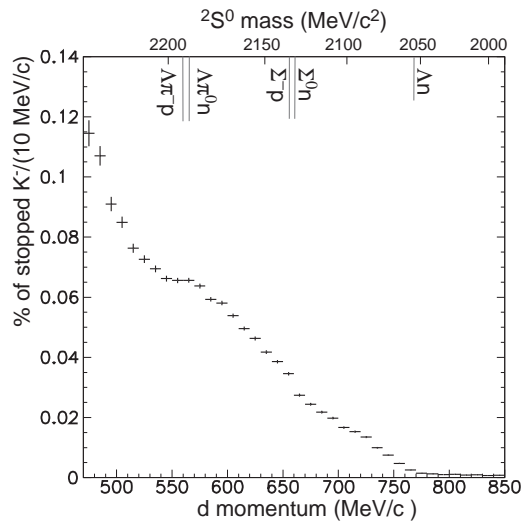}
\caption{\label{fig:mom} A deuteron inclusive momentum spectrum (with statistical errors). Deuterons above 450 MeV/$c$ momentum were detected.}
\end{figure}

The inclusive deuteron momentum spectrum presented as percent fraction of the stopped kaons and per 10 MeV/c momentum is shown in Fig.~\ref{fig:mom}.  The normalization can be performed with the measured free decay ratio of stopped $K^-$ on $^4$He. As a result, a high momentum deuteron component, which originates from non-mesonic reactions, is observed as a $\sim$1$\%$ fraction of stopped $K^-$. An experimentally identified non-mesonic reaction channel producing deuterons is a $\Lambda n$ branch of the 2NA process, $K^- ``pn"(d) \rightarrow \Lambda n (d)$,
where $d$ is produced as a spectator.  In this case, which is known to have the reaction rate of 3.5$\pm 0.2\%$~\cite{ROOSEN}, the deuteron has a momentum up to $\sim$250 MeV/$c$. Thus, the high-momentum component 
over 500 MeV/$c$
with such a large fraction as presented here is quite unexpected, and  it clearly shows evidence for the presence of unknown non-mesonic reactions in which the deuteron appearing in the final state is not a spectator. Since the high-momentum deuterons could be candidates either for the formation or decay signals of ${}^2$S$^0$ or ${}^3$S$^+$ strange multibaryons, or 3 or 4NA processes, it is extremely interesting to study further their origin by correlation studies between $d$ and other particles, such as $\Lambda$ and $\Sigma$, detected in coincidence. In the following, we present the result of the 
$\Lambda d$ correlation analysis. 

\begin{figure*}
\includegraphics[scale=.25]{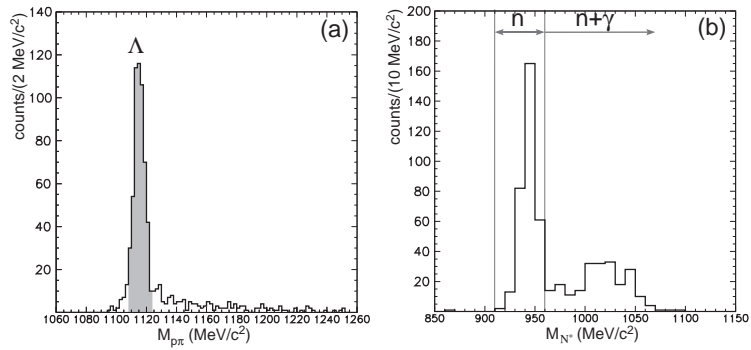}
\caption{\label{fig:pid2} (a) The $M_{p\pi}$ distribution for the events in which back-to-back $p$ and $d$, and additional $\pi^{\pm}$
%
 %
 %
 emitted vertically
  are observed in coincidence. The shaded area is defined as $\Lambda$. (b) The $M_{N^*}$ spectrum, in which the $\Lambda d n$ and $\Sigma^0(\Lambda\gamma) d n$ final states are clearly separated.}
\end{figure*}

 We identified 543 
  $\Lambda d$ pairs from 
  %
  back-to-back $dp$ and vertical-$\pi^{\pm}$-coincidence events. 
  As shown in Fig. \ref{fig:pid2} (a), the peak position of the $p\pi$ invariant mass ($M_{p\pi}$) agrees with the known $\Lambda$ mass, and the width of the $\Lambda$ peak is as narrow as 10 MeV/$c^2$ FWHM, which is fully consistent with the observed $1/\beta$ resolution on each detector arm. 
 By the measurement, the angular region of $-1 \le \cos{\theta_{\Lambda d}} \le -0.6$ was covered, where $\theta_{\Lambda d}$ is the opening angle between $\Lambda$ and $d$ 3-momenta in the laboratory frame, and thus observed $\Lambda d$ pairs are back-to-back correlated.
 Because of the limited acceptance
 of their momenta,
 only energetic $d$ and $\Lambda$ were detected. Therefore, they are considered to be  mainly produced in non-mesonic final states,  
\begin{eqnarray}
(K^- {}^4\textrm{He})_{\textrm{atomic}} &\rightarrow& \Lambda+d+n, \\
                                                                     &\rightarrow& \Sigma^0(\Lambda \gamma)+d+n.
\end{eqnarray}
The $\Lambda d n$ and $\Sigma^0 d n$ final states are separated from possible contaminants, such as $\Lambda/\Sigma^0 \pi dN$, by reconstructing the missing mass
\begin{equation}
M_{N^*}=\sqrt{(p_{init}-p_{\Lambda}-p_{d})^2},
\end{equation}
where $p_{init}$, $p_{\Lambda}$, and $p_{d}$ are 4-momenta of the initial state $K^-$+$^4$He at rest,  and the measured ones of $\Lambda$ and $d$, respectively. The distribution of the thus determined missing mass $M_{N^*}$ is shown in Fig. \ref{fig:pid2} (b). The narrow peak structure at $\sim 940$ MeV/$c^2$ is due to a $\Lambda dn$ final state, whereas a $\Sigma^0(\Lambda \gamma) dn$ final state causes the broad distribution peaked at $\sim 1020$ MeV/$c^2$. As expected, no event exists above $m_{\pi}+m_N \approx$1080 MeV/$c^2$, where $\Lambda \pi dN$ and $\Sigma^0 \pi dN$ final states should appear. Therefore, we selected the $\Lambda dn$ final state by the condition $920 \le M_{N^*}\le 960$ (MeV/$c^2$). 

\begin{figure}
\includegraphics[scale=.35]{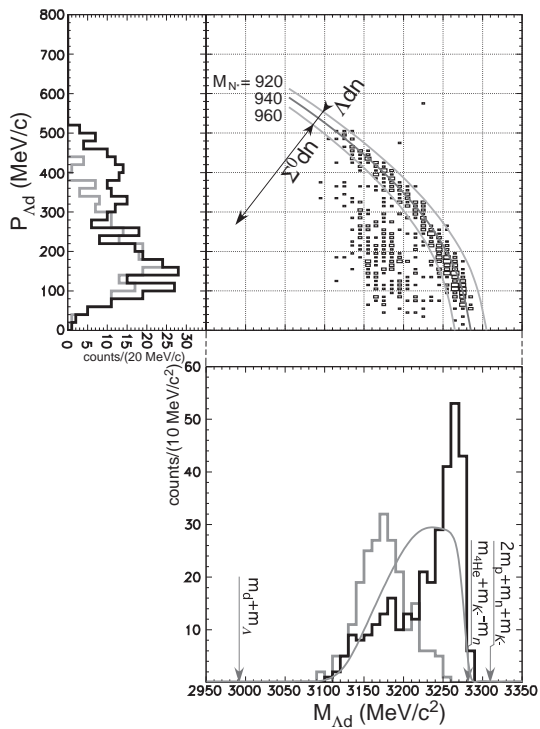}
\caption{\label{fig:inv} A correlation diagram between the $M_{\Lambda d}$ and $P_{\Lambda d}$, with kinematical constraints for $M_{N^*}=$920, 940, and 960 MeV/$c^2$ overlaid. On the projections, contributions of the $\Lambda d n$ and $\Sigma^0 d n$ events are represented by black and gray lines, respectively. The phase space distribution is represented by a thin gray curve on the $M_{\Lambda d}$ spectrum.}
\end{figure}

The correlation between the $\Lambda d$ invariant mass ($M_{\Lambda d}$) and the total 3-momentum ($P_{\Lambda d}$) from all $\Lambda d$ events is shown in Fig. \ref{fig:inv}, where its projections onto the horizontal and vertical axes classified by the $\Lambda dn$ and $\Sigma^0dn$ final states are shown together. 
%
A simulated shape, evaluated 
by uniformly generated $\Lambda d n$ events in the three-body phase space, taking the realistic experimental setup into account, is overlaid on the $M_{\Lambda d}$ spectrum, normalized to the observed number of $\Lambda d n$ events. The $M_{\Lambda d}$ spectrum of $\Lambda d n$ events, which clearly deviates from the simulated one, consists of two components. One is an asymmetric peak located just below the $m_{^4\textrm{He}}+m_{K^-}-m_n$ mass threshold at 3282 MeV/$c^2$, and the other is a broad component from 3100 to $\sim 3220$ MeV/$c^2$. The $M_{\Lambda d}$ resolution near the threshold, estimated from the observed $M_{N^*}$ distribution, is $\sim 8$MeV/$c^2$ rms, which is significantly smaller than the observed width of the peak structure. Identifying $P_{\Lambda d}$ as the momentum of missing neutron, the high-mass peak is correlated with neutrons in the momentum range $\lesssim250$ MeV/$c$. Thus, we can interpret this peak as the $\Lambda d$ branch of the 3NA process,  
\begin{equation}
K^-``ppn"(n) \rightarrow \Lambda d(n),
\end{equation}
where the missing $n$ is a spectator of the reaction, inheriting its original Fermi momentum distribution from $^4$He. The deuteron in the final state could be either from an original $d$ cluster in ${}^4$He participating in the reaction ($``ppn"$ is actually $``pd"$, then), or a product of coalescence after the absorption. The nature of the broad lower mass component accompanying neutrons with momenta higher than $\sim$250 MeV/$c$ is very interesting but still unclear at this moment, and several explanations may be possible.

\begin{figure}
\includegraphics[scale=.40]{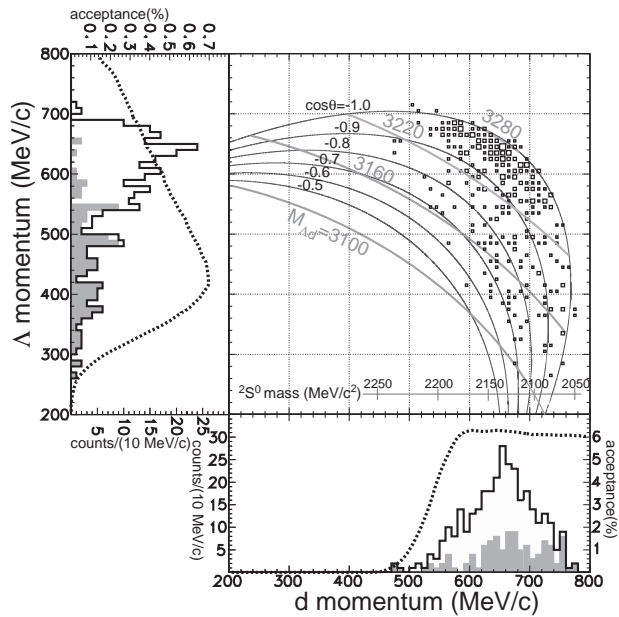}
\caption{\label{fig:ldmom_2d} A correlation diagram between $d$ and $\Lambda$ momenta of $\Lambda dn$ events. 
 Kinematical constraints for $M_{\Lambda d}=const.$ and $\cos{\theta_{\Lambda d}}=const.$ are shown on the correlation as gray curved line segments and black curves, respectively. 
 The projections are also exhibited, where the shaded area expresses the component with $M_{\Lambda d} \le 3220$ MeV/$c^2$. The momentum acceptance, calculated for inclusive $d$ and $\Lambda$ events, are shown by 
 %
 broken
 curves.
}
\end{figure}

The correlation between the momenta of the $\Lambda$ and $d$ of the $\Lambda d n$ events is shown in Fig. \ref{fig:ldmom_2d}. Well-correlated high-momentum $\Lambda d$ pairs constitute the 3NA component
at the region of $\cos{\theta_{\Lambda d}} < -0.9$, 
in which the momenta of $d$ and $\Lambda$ widely distribute along kinematically allowed curves for given $M_{\Lambda d}$ values, reflecting the original Fermi motion. On the other hand, the lower invariant mass component is composed of relatively slow-$\Lambda$ and fast-$d$ pairs, significantly different from the 3NA component. A presumable interpretation of the observed lower mass distribution with conventional processes might be a sequence of a $\Sigma n$ branch of 2NA process and successive $\Sigma\Lambda$ conversion, 
\begin{equation}
K^-``NN"(NN) \rightarrow \Sigma n (NN), \thickspace \Sigma (NN) \rightarrow \Lambda d.
\end{equation}
There are other possible candidates for conventional explanations with two-step reaction mechanisms. One possible exotic interpretation of the lower mass component is to assume the ${}^3$S$^+_{T=0}$ production and its decay to $\Lambda d$. Another possibility  is the ${^2}$S$^0_{T=1/2}$ production and its decay to $\Lambda n$. For both, the widths are expected to be rather broad,  and hence their binding energies and widths are difficult to determine using the present kinematically restricted data. Such interpretations will be examined in detail with full Monte-Carlo simulations in forthcoming publications.

 The observed $M_{\Lambda d}$ distribution of the $\Sigma^0 dn$ events is well interpreted as a $\Sigma^0 d$ branch of the three-nucleon absorption and successive electromagnetic decay of the $\Sigma^0$,
\begin{equation}
K^-``ppn"(n) \rightarrow \Sigma^0 d(n),\thickspace \Sigma^0\rightarrow \Lambda (\gamma),
\end{equation}
where a spectator $n$ and a $\gamma$ from the decay of the $\Sigma^0$ are missing. In this case, the observed $M_{\Lambda d}$ spectrum for $\Sigma^0 d n$ events is translated into a $M_{\Sigma^0 d}$ spectrum by shifting $M_{\Lambda d}$ by about 74 MeV/$c^2$ upward. Thus, we find that the obtained $M_{\Sigma^0 d}$ shape is similar to the asymmetric peak structure of the $M_{\Lambda d}$ spectrum of the $\Lambda d n$ events. This fact implies that there exist nearly the same peak structures just below the threshold on both $T=0$ and $T=1$ $Y^0 d$ channels. 
The interpretation of 
%
these two asymmetric peaks in $M_{\Lambda d}$ in terms of bound-state formation,
as proposed in ~\cite{Piano}
for a similar peak in $M_{\Lambda d}$ from $^6$Li,
 would mean that there are  two distinct species of ${}^3$S$^+_{T=0,1}$ with quite similar masses and widths. This is unlikely, while the 3NA interpretation gives a consistent understanding of both spectra. 
Unfortunately, our $\Lambda d$ measurement has no acceptance to a low-mass component of $M_{\Sigma^0 d}$.

In summary, we have investigated high-momentum deuterons participating in non-mesonic reactions, and observed at least 2 components in the $\Lambda dn$ final state. One consists of well-correlated fast $\Lambda d$ pairs, which is evidence for a $\Lambda d$ branch of the 3NA process in $^4$He. The other is made of pairs of slow $\Lambda$ and fast $d$, which could be interpreted either as coming from exotic signals of ${}^2$S$^0$/${}^3$S$^+$ or as two-step reaction schemes of conventional processes. The observed $M_{\Sigma^0 d}$ spectrum of the $\Sigma^0 dn$ events is well understood as a $\Sigma^0 d$ branch of the 3NA process. 

We are grateful to the KEK staff members of the beam channel group, accelerator group, and electronics group, for support of the present experiment. We also owe much to T. Taniguchi and M. Sekimoto for their continuous contribution and advices for electronics. This work was supported by KEK, RIKEN, and Grant-in-Aid for Scientific Research (S) 14102005 of the Ministry of Education, Culture, Sports, Science and Technology of Japan.

\bibliography{ldcorr2}

\end{document}